\documentclass[aps,preprint,a4paper,12pt]{revtex4-1}
\usepackage[utf8]{inputenc}
\usepackage{amsmath}
\usepackage{hyperref}
\usepackage{amsfonts}
\usepackage{amssymb}
\usepackage{graphicx}
\usepackage{latexsym}
\usepackage{color}
\usepackage{booktabs}
\usepackage{dcolumn}
\usepackage{epsfig}
\usepackage{subfigure}
\usepackage{float}
\usepackage{multirow}
\usepackage{ulem}
\usepackage{xcolor}
\usepackage{svg}
\usepackage[mathcal]{euscript}
\usepackage{tikz}
\def\be{\begin{eqnarray}}
\def\ee{\end{eqnarray}}
\begin{document}

\title{Quasinormal modes of charged BTZ black holes}

\author{R. D. B. Fontana}
\email{rodrigo.dalbosco@ufrgs.br}
\affiliation{Universidade Federal do Rio Grande do Sul, Campus Tramandaí-RS
\\
Estrada Tramandaí-Osório, CEP 95590-000, RS, Brazil and \\
Departamento de Matemática, Universidade de Aveiro, \\
Center for Research and Development in Mathematics and Applications (CIDMA), Campus Santiago, 3810-183, Aveiro, Portugal}

\date{\today}

\begin{abstract}

We investigate the scalar field equation in a (2+1)-dimensional charged BTZ black hole. The quasinormal spectra of the solution are obtained applying two different methods and good convergence between both is achieved. Using the characteristic integration technique we tested the geometry evidencing its stability against linear scalar perturbations. As a consequence a two pattern set of frequencies (families) emerges, one oscillatory and another purely imaginary. In that spectra, the fundamental modes without angular momentum are hugely affected by the presence of the black hole charge surprisedly even for small values of this. Their evolution are controlled by purely imaginary frequencies as in the non-rotating chargeless BTZ case. Similar to others AdS black holes, the fundamental oscillations scale the black hole event horizon and the temperature of the hole (far from maximal charges).

\end{abstract}

%\begin{keyword}
%\texttt{elsarticle.cls}\sep \LaTeX\sep Elsevier \sep template
%\MSC[2010] 00-01\sep  99-00
%\end{keyword}

%\end{frontmatter}

\maketitle

%----------------------------------------------------------------------------------------------------%
\section{Introduction}\label{sec1}

Black holes in three dimensional gravity is an extensive area of research of the past decades. Since the pioneer works of Bañados, Teitelboim and Zanelli \cite{Banados:1992wn}, Jackiw \cite{jackiw1} and Mann \cite{mana1} hundreds of studies were realized analyzing different aspects of those geometries. Particularly, the BTZ solution with or without rotation represents a simple and charming geometry used as toy model for gravity (for a recent thorough study of the "zoo" of solutions for the $(2+1)-D$ Einstein-Maxwell equations refer to \cite{Podolsk__2022}). The spacetime is Anti-de Sitter in its boundary thus possessing interesting conformal structure for AdS/CFT exploration \footnote{In the absence of mass a recent work correlated the spacetime with properties of the graphene \cite{iorio1}}. Also, the thermodynamical properties (e. g. temperature and entropy) have similar derivation as their (3+1)D black holes companions, connected with dynamical features of the spacetime.  Particularly, the introduction of charge into the solution brings a logarithm term to the metric with robust consequences \cite{Carlip_1995,Martinez_2000}. Namely the existence of horizons is ruled by different limitations between the geometric quantities of the geometry (mass, charge and angular momentum).  

The charged BTZ black hole (CBTZ) was studied in several different papers in the last years. The formation of horizons together with the cosmic censorship hypothesis were tested in \cite{Song_2018} and in \cite{Duzta__2020} (semiclassical contemplated e. g.). The duals of the field theory following the AdS/CFT correspondence were investigated in \cite{Maity_2010} and the thermodynamical aspects of the black hole as the Hawking radiation and other semiclassical aspects were considered in \cite{Li_2007,Jiang_2007,Ejaz_2013, Hortacsu_2003}. The geodesical motion of free particles was examined in \cite{Soroushfar_2016} (including the non-linear theory) and an interesting work on the dynamical evolution of scalar shells was performed in \cite{shar1}. From the point of view of the formation of such singular structures (i. e. their existence in the theory) the collapse in lower dimensional gravity was studied in \cite{1702.04601, 1404.0589, 1309.1629, 0008060}.

Similar solutions to the simplest BTZ black hole in $(2+1)$ dimensions were recently discovered - those are characterized by additional geometry properties like acceleration \cite{Astorino_2011} or different global structure (i. e. the non-linear theory \cite{mazha} and quasi-topological gravity \cite{Bueno_2021}). Their scalar perturbation are considered in \cite{Gonzalez_2021, Aragon_2021} and \cite{fon23}. 

One important outcome of such perturbations is the quasinormal spectrum of the black hole, represented as sets of complex numbers $\omega = \Re (\omega )+ \Im (\omega )i$ that encodes information e. g. about the stability and relaxation time of the geometry. In AdS/CFT correspondence, the quasinormal frequencies play also an interesting role, being directly connected to the field theory in smaller dimension: the poles of the two point correlation function in a $(1+1)$ conformal field theory correspond to the frequencies of the rotating BTZ black hole in $(2+1)$ dimensions \cite{Birmingham_2002}. Still more unsettling, instabilities delivered by adding field perturbations in the gravitational system are associated with the phase transition of conductivity in the field theory \cite{Hartnoll_2008, Chen_2012}. 

The quasinormal frequencies in BTZ black holes were widely studied in the past years \cite{2209.01798, 2205.15610, 2005.04084, 1810.08991, 1608.05299VI1, 1112.4619, 1109.5221, 1001.5106, 0908.2657, 0803.0604, 0605027, 0605022, 0504175, 0411267, 0305113, 0402048, 0101194, Cardoso_2001} with many different purposes. They were analytically calculated for the first time in the non-rotating black hole in \cite{Cardoso_2001}. The modes of the rotating geometry were also analytically determined in \cite{0101194}. The missing frequencies of the charged black is one of the aims of this work.

The vast references of the (2+1) black hole geometry when perturbed by probe fields allows us to emphasize three interesting features: the relation of the perturbations to the area spectra of the black hole \cite{1001.5106}, the influence of a non-minimally coupling with the field and its consequences to the vibrations \cite{2005.04084, 1810.08991} and the Choptuik scaling presented in \cite{0101194}.

For black holes of (2+1)-dimensional alternative theories, the frequencies were also extensively investigated since the 90's \cite{1906.06654, 1906.04360, 1904.10847, 1901.00448, 1801.02555, 1801.03248, 1711.04146, 1404.5371, 1407.6394, 1003.1381, 0908.0057, 0903.1537, 0903.0088, 0802.3321, 0306214}. Also in lower dimensional gravity (1+1) the presence of the quasinormal modes is noticed for different geometric arrangements \cite{1201.3605, 2011.08179, 0408042}.

When the coupling between the field and the geometry is non-trivial \cite{1308.3076, 1901.07310, 2101.06410, 1810.08822}, 
the existence of scalar hairy configurations \cite{2305.00686, 2305.00058, 1405.2956, 1412.4002, 1408.2419, 0201170, 9805011} or related systems (as e. g. cloud of strings \cite{1810.08822, 1707.08133, 1810.08830}) brings significant modification of the spacetime properties (for charged black holes, see \cite{1305.5446, 1408.1998, 1602.04021}). The first order perturbations of these hairy black holes is still an open branch of research not much investigated to the date that can bring valuable information as dynamical stability limits for the theory couplings.

The stability of black holes geometries to small field perturbations can be investigated jointly with the quasinormal modes query. When studying the field propagation problem and its (possible) decomposition in an two-dimensional wave equation, integration in double null coordinates can generally be applied given rise to a twofold prospect: the stability analysis and the acquisition of the quasinormal frequencies. With those spectra, interesting features of black hole are at hand: the scalings of the modes in AdS spacetimes, breakdown of isospectrality \cite{rod1,rod2} and the emergence of multiple families of solutions. Those prospects are the aim of the present work.

The article is organized as follows, in section \ref{sec2} we describe the charged BTZ black hole with its main geometrical features and in scetion \ref{sec3} we review the scalar field equation and the necessary numerics to study the quasinormal problem. Section \ref{sec4} is devoted to our results and analysis followed by the discussions on section \ref{sec5}. 

%----------------------------------------------------------------------------------------------------%
\section{The Charged BTZ black hole}\label{sec2}

We consider a charged BTZ black hole in $(2+1)$ dimensions with line-element written as \cite{Banados:1992wn,Carlip_1995,Martinez_2000}
\be
\label{lin}
ds^2= - fdt^2 + f^{-1}dR^2 + R^2dx^2, 
\ee
in which
\be
\label{lap}
f= -m+\frac{R^2}{L^2}-\frac{q^2}{2}\ln \left[ \frac{R}{R_0} \right] \equiv -M +r^2- Q^2 \ln (r).
\ee
Here $f$ is the lapse function with two real solutions $f=0$, the Cauchy ($r_c$)  and event ($r_h$) horizons, provided that $M \geq 1$. In this case, a black hole with two horizons is formed in the geometry and cosmic censorship in all versions is preserved outside the event horizon. 

The metric (\ref{lin}) with the first form of the lapse function in (\ref{lap}) appears e. g. in \cite{Carlip_1995, 0110289, 1208.6251, 1608.05299VI1, kone}, with the proper adjustment of its constants. The second form of (\ref{lap}) can be found e. g. in \cite{Martinez_2000} (Eq. 28) with the proper adjustment of $Q$. Here, for simplicity, we redefine the $R$ coordinate in terms of $R=rL$ to accommodate the cosmological radius outside the wave equation. The consequences of a different cosmological radius to the scalar perturbation is commented in the last paragraph of section IV.

For a particular value of $r_h$, $Q_{max} = \sqrt{2r_h}$ represents the maximum charge. Such value however is not extremal, since adding a small amount $\delta Q$ to the geometry will not destroy the horizons, but increase the black hole size ($r_h$). On the other hand for small mass terms ($M \leq 1$), the subtraction of small amounts of charge may lead to a naked singularity.

%The spacetime is static with two Killing vectors, $\partial_t$ and $\partial_x$.

The existence of two horizons for the geometry imposes restrictions on the values of $M$ and $Q$. Here, the logarithm term in $f$ brings non trivial causal implications for the limits on $M$ and $Q$, one of them mentioned above. Even more sounding, for certain values of $Q$, $M<0$ allows for solutions of $f=0$ \cite{Martinez_2000}. The issue is circumvented by renormalizing $M$, defining the mass of the black hole as $M_0 = M +Q^2\ln r_+ = r_+^2$, that opportunely represents a BPS-like bound for $M_0$  \cite{Cadoni_2009}. Interestingly enough, the mass term of the black hole imposes a threshold in the charge constant as
\be
\label{esq1}
M_0 \geq Q^2 /2,
\ee
similar to that present in higher dimensional companions.

Since the aim of our investigation is the study of the dynamical stability of the charged black hole to small scalar perturbation and the achievement of the quasinormal modes of such geometry, we restrain our simulations to the scope of $M \geq 1$ where two horizons are always present. 

We consider the propagation of a probe massless scalar field as a small perturbation in the geometry introduced via
\be
\label{act}
\mathcal{S}_m = \int_{\cal M} d^3y \sqrt{-g}\partial_\mu \Phi \partial^\mu \Phi. 
\ee
The variation of this action produces the motion equation, 
\be
\label{sfe0}
\Box \Phi =0
\ee
that we explore in the next section.

\section{Motion equation and numerics}\label{sec3}

\begin{figure}
\begin{center}
\includegraphics[width=0.45\textwidth]{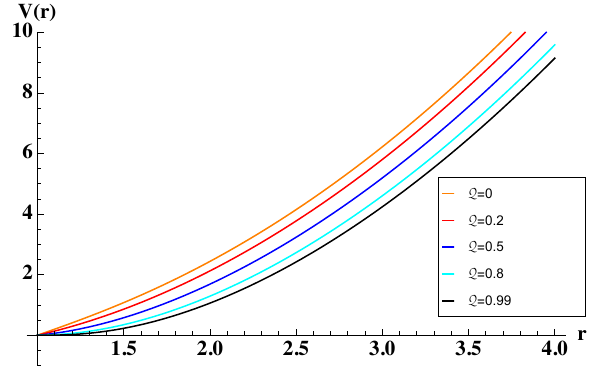}
\includegraphics[width=0.45\textwidth]{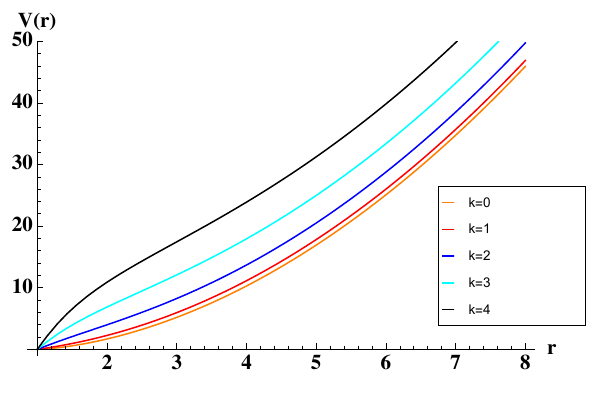}
\end{center}
\caption{The scalar field effective potential in a charged BTZ black hole with $r_h = 1$ and $k=0$ (left), $\mathcal{Q}=0.5$ (right).}
\label{fg0}
\end{figure}

A scalar field acting in the background of the CBTZ black hole has its evolution dictated by the above Klein Gordon equation also written in the form,
\be
\label{sfe}
\frac{1}{\sqrt{-q}}\partial_\mu ( \sqrt{-g} g^{\mu\nu} \partial^\nu \Phi ) = 0.
\ee
Choosing an usual Ansatz for the scalar field decomposition in $(2+1)-D$,
\be
\label{ans}
\Phi = \frac{\psi (r,t)}{\sqrt{r}} \chi (x) =\frac{\Psi (u,v)}{\sqrt{r}} e^{ik x}
\ee 
we apply the metric (\ref{lin}) of the CBTZ into (\ref{sfe}), which, in the double null coordinate system $u$ and $v$\footnote{Those are conventionally defined as $du = dt-dr_*$ and $dv = dt+dr_*$ in which $dr_* = f^{-1}dr$ is the customary tortoise radial coordinate.} turns this equation to 
\be
\label{sfe2}
\left( -4 \frac{\partial}{\partial u}\frac{\partial}{\partial v} + V(r) \right) \Psi (u,v) =0, 
\ee
in which $V(r)$ represents the scalar field potential given by
\be
\label{pot}
V(r) = f\left( \frac{f}{4r^2}-\frac{1}{2r}\frac{\partial f}{\partial r} - \frac{k^2}{r^2}\right).
\ee
In figure \ref{fg0}, we provide different plots representing the potential with multiple geometric and field parameters.

Now, to obtain the quasinormal modes we employ the ordinary characteristic integration (by means of which the wave signal is achieved) together with the prony method for filtering the frequencies values. (Purely imaginary solutions may also be analyzed by simpler linear regression with the extraction of its damping). This and the Frobenius expansion (utilized as a control method) are briefly described in what follows. 

%---------------
%\begin{figure}[htb]
%\begin{center}
%\includegraphics[height=10.5cm, width=10cm]{diagrama_bh_singular_regular.pdf}
%\caption{({\it{Left diagram:}}) Conformal diagram for maximal extended black hole %solution spacetime $\alpha_2\neq 0$. ({\it{Right diagram:}})  Conformal diagram for %regular black hole with one event horizon.The blue lines represent the surfaces in which %the radial coordinate is constant}
%\label{diagrams}
%\end{center}
%\end{figure}
%---------------

%----------------------------------------------------------------------------------------------------%
\subsection{Methods}

The characteristic integration in double null coordinates was firstly developed in \cite{Gundlach_1994}. It consists in the discretization of the null coordinates $u$ and $v$ and the choice of acceptable physical data  for an evolution schematic diagram $u$ x $v$ \footnote{A detailed description can be found in \cite{Konoplya:2011qq}.}. In AdS spacetimes such discretized equation, derived from (\ref{sfe2}) can be put into the form
\be
\label{dfe}
\Psi_N = \left(1+ \frac{h^2}{16}V_S \right)^{-1}\left( \Psi_E + \Psi_W - \Psi_S -\frac{h^2}{16}(\Psi_N V_N + \Psi_S V_S + \Psi_E V_E)\right).
\ee
Initial surfaces that are physically relevants are those coherent with boundary conditions. In AdS cases, we may take 
\be
\label{ida}
\Psi (u,v=v_0) = gaussian \hspace{0.2cm} package, \hspace{0.5cm} \Psi (r\rightarrow \infty ) = 0 
\ee
The quasinormal signal represents part of any surface $\Psi (r=r_0 , t)$ collected in the grid of integration after the evolution scheme is completed (preferable far from the asymptotic region, where the above second boundary condition is implemented). Such surface constitutes a matrix (vector) of data used to extract the frequencies, supposing a decomposition of type
\be
\label{pro}
\Psi = \sum_n A_n e^{-i \omega_n t}.
\ee
All quasinormal modes $\omega_n$ are encapsulated in this field decomposition, which is expected for the numerical results of field integration. In order to obtain the fundamental and excited modes, a particular even number of modes $n$ must be chosen in (\ref{pro}), though the higher the overtone the trickier its calculation (being the fundamental the easiest mode). 

An important step in this calculation is the establishment of the tortoise as a function of $r$, what can not be done analytically. In this case we use a two step approximation, expanding the integral near the points $r=r_h$ and $r \rightarrow \infty$, subsequently matching the resultant expressions. A more thorough description of such procedure is introduced in appendix \ref{app1}.

The second method we use to compare the data obtained in numerical integration is the Frobenius expansion, similar to that developed in \cite{hor2000}. In order to implement it, we consider the line-element (\ref{lin}) in coordinates $v, r$ and $x$, applying it to the Klein-Gordon equation (\ref{sfe}) with the ansatz (\ref{ans}). Now the variable transformation $z= 1/r$ allows us to write the field equation in the form
\be
\label{sfe3}
z^4 F \ddot{\Psi} + (2z^3 F + z^4 \dot{F} + 2i\omega z^2)\dot{\Psi} + \frac{1}{4}(2z^3\dot{F}+z^2 F -4z^2k^2)\Psi = 0,
\ee
in which the dot represents a derivative with relation to $z$ and $F = f|_{r\rightarrow 1/z}$. By defining three new functions 
\be
s(z) & = & z^4 F,  \\ 
\tau (z) & = & 2z^3 F + z^4 \dot{F} + 2i\omega z^2, \\
u(z) & = & \frac{1}{4}(2z^3\dot{F}+z^2 F -4z^2k^2),
\ee
we expand each one around $z_h$ ($=r_h^{-1}$) i. e., 
\be
s(z) & = & \sum_{n=0}s_n (z-z_h)^n,  \\ 
\tau (z) & = & \sum_{n=0}\tau_n (z-z_h)^n, \\
u(z) & = & \sum_{n=0}u_n (z-z_h)^n.
\ee
Now, implementing $\Psi \rightarrow \sum_n (z-z_h)^{n+\alpha}$ in the field equation we can adjust the boundary condition of ingoing wave at the horizon. To leading order in $z-z_h$, the solutions for the indicial equation are $\alpha =0$ (ingoing) and $\alpha = 1- \frac{\tau_0}{s_1} = \frac{2i \omega}{\dot{F}}\Bigg|_{z=z_h}$ (outgoing - to be avoided). 

In a next stage, the ansatz must be substituted in the motion equation yielding the recurrence relation
\be
\label{rre}
a_n = -\frac{1}{(n-1)ns_1 + n\tau_0} \sum_{j=0}^{n-1} \Big( (j-1)j\hspace{0.03cm} s_{n-j+1}+j\hspace{0.03cm} \tau_{n-j}+u_{n-j-1} \Big) a_j.
\ee

The last step is the implementation of the second boundary condition, $\Psi|_{z=0} = \sum_n a_n (-z_h)^n =0 $ in a computational level. To do that, the series must be truncated at a specific number of terms and the convergence can be tested increasing this number. 

%-------- --------------------------------------------------------------------------------------------%
\section{Results}\label{sec4}

The fundamental quasinormal modes ($n=0$) calculated for black holes with different $r_h$ and charges obtained through characteristic integration are listed in table \ref{t1}. The results displayed were checked with the Frobenius method until $Q\leq r_h$. A comparative table between both methods is given in the appendix \ref{app2}.

\begin{table}[h]
  \centering
 \caption{{\color{black} Quasinormal modes of the massless scalar field with $k=0$. For simplicity we use a fraction of $Q$ relative to the maximum charge $Q_{max}$ defined as $\mathcal{Q}= Q^2/Q_{max}^2$. All frequencies are purely imaginary and stable, $\Im (\omega ) < 0$}.}
\addtolength\tabcolsep{6pt}
    {\color{black}\begin{tabular}{c|ccc}
    \hline    \hline
$\mathcal{Q}$ & $r_h=1$ & $r_h=10$ & $r_h=100$ \\
%    \hline
	 &	\multicolumn{3}{c}{$-\Im (\omega )$}\\
\hline \hline
0	&	1.99558	&	19.9529	&	199.509	\\
0.1	&	1.44617	&	14.4615	&	144.615	\\
0.2	&	1.17840	&	11.7839	&	117.840	\\
0.3	&	0.96541	&	9.65401	&	96.5430	\\
0.4	&	0.78290	&	7.82894	&	78.2983	\\
0.5	&	0.62109	&	6.21092	&	62.1365	\\
0.6	&	0.47488	&	4.74882	&	47.5718	\\
0.7	&	0.34125	&	3.41243	&	34.1340	\\
0.8	&	0.21819	&	2.18312	&	21.8312	\\
0.9	&	0.10463	&	1.04612	&	10.4612	\\
 \hline  
    \end{tabular}}
  \label{t1}
\end{table}
The first value of table \ref{t1} represents the BTZ limit ca. $0.2\%$ deviant from the analytical result, $\omega = k-2r_h i$ \cite{Cardoso_2001}. 

In table \ref{t1} we observe the sensitive variation of the modes with the accretion of charge: even small black hole charge variations affect strongly the quasinormal frequencies. We can see that such is not the case in higher-dimensional black holes: small variations of charge (far from the extremal limit) produces similar small variations in the quasinormal spectra. For example, the fundamental mode of the Reissner-Nordström-Anti-de Sitter (RNAdS) metric, as in figure 4 of \cite{Molina_2004} (in the small charge regime) varies very little with increasing $Q$. That sort of behavior is present also in the Reissner-Nordström-de Sitter (RNdS) geometry \cite{Wang_2004} and in the Reissner-Nordström (RN) black hole where such influence is even smaller \cite{Konoplya_2002}. On the other hand, at intermediate charges in RNAdS (or near extremal charges for RN and RNdS \cite{Cardoso_2018}), the purely imaginary oscillations are dominant. In that regime, small variations of the black hole charge affect $\omega$ strongly. The difference is crucial: it reveals the presence of two families of frequencies dominating the spectrum for different charge regimes in all three spacetimes (4-dimensional).

The results of the charged BTZ metric (\ref{t1}), on the other hand, bring only non-oscillatory fundamental frequencies that are highly influenced by the change in $\mathcal{Q}$. Such outcome manifest the absence of an oscillatory behavior for the scalar field at non-excited levels, very particular from BTZ black holes, a profile that, as mentioned above, appears as dominant only for intermediate or near-extremal charge regimes in RNAdS and RN/RNdS geometries respectively.

%It can be observed The novelty of the presented frequencies in table \ref{t1} is the significant decrease of the damping with increasing charge: even small charges affect strongly the quasinormal frequencies. That is not the case when we compare such scalar oscillations with those of the Reissner-Nordström-Anti-de Sitter solution: the fundamental mode varies in a few percent for small increments of the charge \cite{Wang_2004, Molina_2004}

% The reason for such difference is the non-oscillatory character of the dominant family of modes present in the CBTZ black hole. As we can see v. g. in figure 9 of \cite{Wang_2004} the oscillatory modes (OM) vary slightly with the change in the black hole charge while the non-oscillatory (NOMs) are much more sensitive to it (interestingly these variations of OMs and NOMs occur in opposite directions for increasing charge). 

In the same table \ref{t1}, the relation between $r_h$ and $\omega$ is evident, for each different value of charge ($-\Im (\omega ) \propto r_h$). By inspection a logarithm scale between $\mathcal{Q}$ and $\omega$ of type 
\be
\label{sca1}
\ln \left( \frac{\omega}{r_h}\right) \propto -\mathcal{Q}
\ee
 can be settled (with $R^2\sim -0.98$). Such approximation is however only trustworthy far from the maximal charge limit where, for instance $R^2\sim -0.997$ with $Q\leq r_h$.

\begin{table}%[h]
  \centering
 \caption{{\color{black} Fundamental quasinormal modes of the massless scalar field with $r_h=1$.}}
\addtolength\tabcolsep{6pt}
    {\color{black}\begin{tabular}{c|cccccccc}
    \hline    \hline
$\mathcal{Q}$ & \multicolumn{2}{c}{$k=1$} & \multicolumn{2}{c}{$k=2$} & \multicolumn{2}{c}{$k=3$} & \multicolumn{2}{c}{$k=4$}  \\
%    \hline
$\mathcal{Q}$	 & $\Re (\omega )$ & $-\Im (\omega )$	 & $\Re (\omega )$ & $-\Im (\omega )$	 & $\Re (\omega )$ & $-\Im (\omega )$	 & $\Re (\omega )$ & $-\Im (\omega )$ \\
\hline \hline
0	&	0.99998	&	1.99998	&	1.99998	&	1.99998	&	2.99998	&	1.99998	&	4.00000	&	1.99998	\\
0.1	&	0.78216	&	1.90896	&	1.81173	&	1.98058	&	2.82251	&	2.02275	&	3.82855	&	2.05222	\\
0.2	&	0.53654	&	1.79487	&	1.60901	&	1.96624	&	2.63787	&	2.05640	&	3.65471	&	2.11689	\\
0.3	&	0.23233	&	1.63733	&	1.38841	&	1.96344	&	2.44816	&	2.10525	&	3.48186	&	2.19614	\\
0.4	&	0	&	1.17838	&	1.15183	&	1.98618	&	2.25908	&	2.17301	&	3.31498	&	229018	\\
0.5	&	0	&	0.88479	&	0	&	1.52640	&	0	&	2.11119	&	3.15872	&	2.39622	\\
0.6	&	0	&	0.65900	&	0	&	1.06899	&	0	&	1.47212	&	0	&	1.87393	\\
0.7	&	0	&	0.46685	&	0	&	0.72717	&	0	&	0.99438	&	0	&	1.26414	\\
0.8	&	0	&	0.29649	&	0	&	0.44928	&	0	&	0.60962	&	0	&	0.77211	\\
0.9	&	0	&	0.28405	&	0	&	0.42201	&	0	&	0.56866	&	0	&	0.71782	\\
 \hline  
    \end{tabular}}
  \label{t2}
\end{table}

Now, regarding the AdS/CFT correspondence and its duals \cite{hor2000}, the scale $\omega \propto T_H$ appears only far from $\mathcal{Q} \sim 1$. Considering the Hawking temperature of the hole \cite{Gwak_2016},
\be
\label{th1}
T_H= \frac{4r_h^2-2Q^2}{8\pi r_h} = \frac{r_h}{2\pi}(1-\mathcal{Q})
\ee
for small $\mathcal{Q}$, we have $1-\mathcal{Q}\sim e^{-\mathcal{Q}}$ what brings $ T_H  \propto \omega$. In such case we recover the interpretation of $\Im (\omega)$ as the inverse of the relaxation time for thermal equilibrium in the associated field theory.

In table \ref{t2} we display the quasinormal frequencies for different angular momenta of the field, with multiple black hole charges. The convergence between both methods for the frequencies is the highest of our calculations, with a maximum deviation of ca. $0.03\%$. 

The fundamental mode ($k=n=0$) is purely imaginary as in other BTZ-like black holes \cite{Cardoso_2001, 0101194}. We can see for the charged BTZ geometry, a scale $\omega \propto r_h$ tipical from AdS geometries.

Comparing our results to that of the chargeless rotating BTZ black hole \cite{0101194}, those modes also attenuate as we increase the value of inner horizon (relative to the event one). However such attenuation is boldly enhanced in our imaginary frequencies, contrary to \cite{0101194}. The reason for that is the scale existent in both geometries: in the rotating black hole the fundamental quasinormal modes have $\Im (\omega) = 2(r_h - r_c)$ while in our case we have $\Im (\omega) \propto 2(r_h - Q^2/2)$. This is in accordance with the enhancement of the attenuation since the factor $Q^2/2 - r_c$ increases as we increase $Q$. In the BTZ geometry without rotation and charge \cite{Cardoso_2001} the fundamental mode scales the event horizon $\Im (\omega) = 2r_h$ which clearly has to be properly modified in our charged spacetime. For small to intermediate charges, the scale introduced in (\ref{sca1}) is a very good approximation with slope $\sim 1$. 

An interesting behavior can be explored in that data: the transition between oscillatory and purely imaginary modes for intermediate values of $\mathcal{Q}$. This change evidences the presence of two different families of modes, what we confirm by calculating the frequencies nearby the transition points. Examining e. g. $k=2$, the non-excited oscillatory mode for $\mathcal{Q}=0.5$ reads $0.920-2.043i$ whose imaginary part is higher than the fundamental mode, $-1.526i$. On the other hand, fo $\mathcal{Q}=0.4$ the first purely imaginary mode to appear is $-2.179i$ which is beaten by the fundamental one $1.152-1.986i$. A similar behavior was calculated for $k=3$ whose complementary modes (not shown in the table) read $-3.085i$ ($\mathcal{Q}=0.4$) and $2.073-2.267i$ ($\mathcal{Q}=0.5$). 

The purely imaginary family is substantiated by an almost constant gap between modes as we increase $k$ (same $\mathcal{Q}$). For this family, a distance $\delta k$ represents a proportional distance $\delta \Im (\omega )$ what is not the case for the oscillatory modes \cite{Destounis_2020,Fontana_2021} . On the subject, we observe the case $k=0$ whose family of purely imaginary modes is unique, not correlated to those purely imaginary modes appearing for $k>0$ \cite{Cardoso_2018}. In our case, that frequencies are descendent of the oscillatory modes presented in table \ref{t2}.

\begin{figure}
\begin{center}
\includegraphics[width=0.45\textwidth]{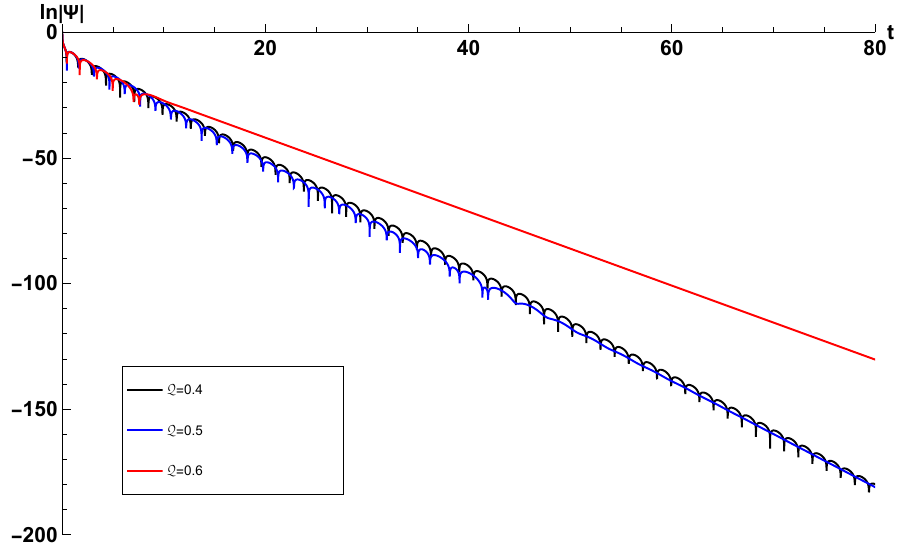}
\includegraphics[width=0.45\textwidth]{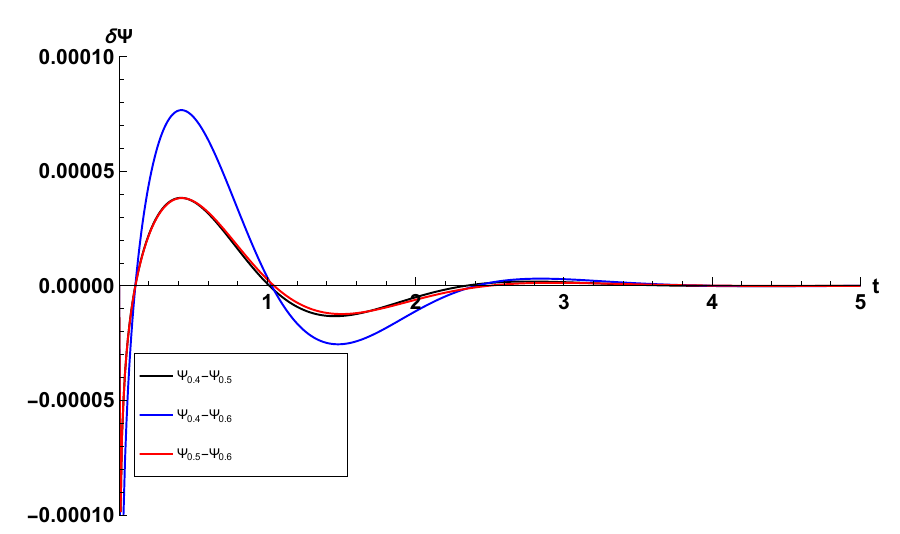}
\end{center}
\caption{The scalar field profile of the charged BTZ black hole with $k = 3r_h = 3$ and the diference in the wave signal for spacetimes with different charges.}
\label{fg1}
\end{figure}

A panel of the transition behavior with three plots is given in figure \ref{fg1}.

The existence of non-oscilatory modes controlling the field evolution is also observed in the non-linear theory with its perturbations. In \cite{Aragon_2021} the authors consider a non-linear Maxwell Lagrangian $(F_{\mu \nu}F^{\mu \nu})^{3/4}$ with the correspondent BTZ-like charged black hole. There, the fundamental modes with $k=0$ follow a similar pattern for an increasing charge (see e. g. Table 1): the attenuation decreases with the increment of $q$. Such attenuation is however, smaller than that of the linear eletrodynamic theory we present in this study (v. g. comparing the results of table 1 in both works). 

We also see a similar qualitative behavior for modes with $k>0$. Up to intermediate black hole charges the scalar field evolution is dominated by oscillatory modes, being the purely imaginary the lowest lying for higher charges. That seems to be the case in such spacetime independent of the logarithm term as in (\ref{lap}).

Considering other modified theories of gravity in high dimensional spaces, such as that of \cite{Zhao_2022, zhao2023quasinormal} we point out two specific differences: in our thorough investigation the effective potential is always positive, not allowing for instabilities in the field evolution. The resulting scalar perturbation evolves consequently into a well-behaved tower of quasinormal modes. We also see (e. g. \ref{t1}) purely imaginary-dominated profiles unnoticed in most of such theories although a more careful examination is needed.

As a last comment, we point up the role of the cosmological constant in the quasinormal frequencies. Considering the metric (\ref{lin}), if we transform the $R$ coordinate according to (\ref{lap}) and the time coordinate as $\tau \rightarrow L^{-1} t$, the scalar motion equation remains the same as that of (\ref{sfe2}). Consequently, the quasinormal frequency is impacted by a factor $L$. That represents finally a rescaling of type $\tilde{\omega}\rightarrow L \omega$ in the modes.

%----------------------------------------------------------------------------------------------------%
\section{Discussion}\label{sec5}

In this work we studied the scalar perturbation of the charged BTZ black hole. The propagation of the scalar field was considered with two different methods - characteristic integration and Frobenius - and the quasinormal modes obtained presenting a very good convergence. The frequencies calculated in the paper exibhit a few interesting features.

First, the geometry was not destabilized in all field profiles calculated in our extensive search through characteristic evolution: the stability of the black hole against scalar field perturbation holds.  

The modes offer two interesting scalings, the usual one proportional to $r_h$ and the other - far from maximum charge values - with the Hawking temperature $\omega \propto T_H$ substantiating the interpretation of $\Im (\omega )$ as the inverse of the time for thermal equilibrium in CFT \cite{hor2000}. 

For higher values of angular momenta, we verify the presence of two families of frequencies fighting to take control of the field evolution. An oscillatory family takes control for small charges whereas a purely imaginary field dominates for higher charges (both families compete in intermediate values, being present in every scenario). By inspecting the BTZ chargeless limit, the oscillatory family is the same present for the special case of $k=0$ but in such, specially represented by purely imaginary evolutions (although further investigation is necessary on the subject). 

Further lines of investigation for future works include the study of the superradiant phenomena \cite{Gonzalez_2021, Ortiz_2012, Ho_1998, Dappiaggi_2018} in the same black hole (that we address in a follow up article \cite{Fontana_2024}) and in charged rotating BTZ black holes (as recently done in \cite{kone} but with an investigation of the modes). Also valuable is the study of the propagation of another probe fields that could impact the geometry differently. 

\section{Acknowledgments}

The author thanks the hospitality of DMAT - University of Aveiro/CIDMA. He is grateful to Carlos A. R. Herdeiro, Clovis A. S. Maia and Jefferson S. E. Portela for fruitful discussions and João E. K. Nauderer for help in running the numerical recipes. The work is supported by the Center for Research and Development in Mathematics and Applications (CIDMA) through the Portuguese Foundation for Science and Technology (FCT— Fundação para a Ciência e a Tecnologia), references https://doi.org/10.54499/UIDB/04106/2020 and https:// doi.org/10.54499/UIDP/04106/2020.

\appendix

\section{Tortoise coordinate integration}\label{app1}

The tortoise coordinate is defined via lapse function as 
\be
\label{tt1}
r_* = \int \frac{1}{f}dr,
\ee
which is not analytical. In order to implement the integration, we solved the above operator in two stages. 

Firstly we define the near-horizon (nh) coordinate $\varepsilon$ as 
\be
\label{tt2}
r = r_h (1 + \varepsilon ) \hspace{0.5cm} \rightarrow \hspace{0.5cm} \varepsilon = \frac{r-r_h}{r_h}.
\ee
In such case (\ref{tt1}) turns to the form
\be
\nonumber
r_*^{(nh)} = \int \frac{r_h}{\varepsilon^2 (r_h^2) + \varepsilon (2r_h^2) -Q^2\ln (1+\varepsilon )}d \varepsilon = \int \frac{r_h}{\sum (B_n+A_n) \varepsilon^n} d \varepsilon = \int \frac{r_h}{\sum b_n \varepsilon^n} d \varepsilon
\\
\label{tt3}
\ee
in which 
\be
\label{tt4}
B_1 = 2r_h^2, \hspace{0.5cm} B_2 = r_h^2, \hspace{0.5cm} B_{n>2} = 0, \hspace{0.5cm} A_n = Q^2\frac{(-1)^{n}}{n}.
\ee
Now, with the new expansion,
\be
\label{tt5}
\frac{1}{\sum b_n \varepsilon^n} = \sum a_n \varepsilon^{n-1}
\ee
we obtain
\be
\label{tt6}
r_*^{(nh)}=r_h \left( a_0\ln (\varepsilon )+ \sum_{n=1}^N a_n \frac{\varepsilon^{n}}{n} \right), \hspace{0.7cm} a_n = -\frac{1}{b_1}\sum_{j=0}^{n-1}b_{n-j+1}a_j
\ee
in which $a_0 = b_1^{-1}$. Such technique circumvents integrability issues for $r_*$ near $r_h$ since it may not be analytically determined, but taken as an approximation by truncating the series.

In the second approximation we expand the tortoise function (\ref{tt1}) in the asymptotic region of the AdS spacetime $(ni)$,
\be
\label{tt7}
r_*^{(ni)} = \int \sum_{i=0}^{N}\sum_{j=0}^{i} \frac{ \mathrm{C}_i^j M^{i-j} Q^{2j} \ln^j (r) }{r^{2i+2}} dr %it would be (r/L)^{2i+2}} dr
\ee
and
\be
\label{tt8}
r_*^{(ni)} = - \sum_{i=0}^{N}\sum_{j=0}^{i}%L^{-2-2i}
(1+2i)^{-1-j}   \mathrm{C}_i^j M^{i-j} Q^{2j} \Gamma [1+j,(1+2i)\ln (r)]. 
\ee
The convergence of (\ref{tt6}) is limited by the maximum value of $r \lesssim 2r_h$, while the convergence of (\ref{tt8}) is limited by a minimum value $r \gtrsim r_h$.
Computationally both approximations fail before those points such that the match between the functions is dictated by the chosen charge: the higher the charge the smaller the convergence of both series. 

Unfortunately this fact prevents the possibility of investigation of geometries with maximum charge and proximal values $Q \lesssim Q_{max}$.

\section{QNM's compared with different methods}\label{app2}

In table \ref{tap2} we compare the quasinormal modes calculated with different methods, characteristic integration (CI) and Frobenius with the Horowitz-Hubeny (HH) scheme \cite{hor2000} truncated at $N=45$ terms. While the Frobenius method converges in a very short time (a few seconds to a few minutes), the matrix of the tortoise coordinate necessary for the CI method is very time consuming: for higher values of charge a complete calculation can endure a few days. We recall that the convergence in the HH method is limited by $Q\leq r_h$, i. e. $\mathcal{Q} \leq 0.5$, while in the CI we have no such limit. (In practice though, it is not possible to go beyond $\mathcal{Q}\sim 0.9$ since the computational time increases exponentially near $\mathcal{Q} = 1$).

\begin{table}[h]
  \centering
 \caption{{\color{black} Comparison of quasinormal modes ($k=0$) obtained with different numerics (in the table, $-\Im (\omega )$). All frequencies are purely imaginary and stable.}}
\addtolength\tabcolsep{6pt}
    {\color{black}\begin{tabular}{c|cccccc}
    \hline    \hline
 %    \hline
$\mathcal{Q}$  &	\multicolumn{2}{c}{$r_h=1$}  &	\multicolumn{2}{c}{$r_h=10$}  &	\multicolumn{2}{c}{$r_h=100$}   \\
 & CI & HH & CI & HH & CI & HH \\
\hline \hline
0	&	1.99558	&	2.00066	&	19.9529	&	20.0057	&	199.509	&	200.057	\\
0.1	&	1.44617	&	1.44630	&	14.4615	&	14.4630	&	144.615	&	144.630	\\
0.2	&	1.17840	&	1.17847	&	11.7839	&	11.7847	&	117.840	&	117.847	\\
0.3	&	0.96541	&	0.96546	&	9.65401	&	9.6545	&	96.5430	&	96.545	\\
0.4	&	0.78290	&	0.78292	&	7.82894	&	7.8292	&	78.2983	&	78.292	\\
0.5	&	0.62109	&	0.62110	&	6.21092	&	6.2110	&	62.1365	&	62.110	\\
 \hline  
    \end{tabular}}
  \label{tap2}
\end{table}

%----------------------------------------------------------------------------------------------------%

%----------------------------%

%\section*{References}

\bibliography{references}

\end{document}